\documentclass[conference]{IEEEtran}
\IEEEoverridecommandlockouts
\usepackage{cite}
\usepackage{amsmath,amssymb,amsfonts}
\usepackage{algorithmic}
\usepackage{graphicx}
\usepackage{textcomp}
\usepackage{xcolor}
\usepackage[T1]{fontenc} 
\def\BibTeX{{\rm B\kern-.05em{\sc i\kern-.025em b}\kern-.08em
    T\kern-.1667em\lower.7ex\hbox{E}\kern-.125emX}}
\usepackage{fancyhdr}

\begin{document}

\title{Ternary circuits: why R = 3  is not the Optimal Radix for Computation}

\author{\IEEEauthorblockN{ Daniel Etiemble}
\IEEEauthorblockA{\textit{Computer Science Laboratory (LRI)} \\
\textit{Paris Sud University}\\
Orsay, France \\
de@lri.fr}

}

\maketitle
\begin{abstract}
A demonstration that e=2.718 rounded to 3  is the best radix for computation is disproved. The MOSFET-like CNTFET technology is used to compare inverters, Nand, adders, multipliers, D Flip-Flops and SRAM cells. The transistor count ratio between ternary and binary circuits is generally greater than the log(3)/log(2) information ratio. The only exceptions concern a circuit approach that combines two circuit drawbacks (an additional power supply and a circuit conflict between transistors) and only when it implements circuits based on the ternary inverter. For arithmetic circuits such as adders and multipliers, the ternary circuits are always outperformed by the binary ones using the same technology.
\end{abstract}

\section{Introduction}\label{sec1}

Multivalued circuits have been studied for more than fifty years. Many ternary and 4-valued circuits have been
proposed using different integrated circuit technologies. Two main arguments have been used to justify these proposals:
\begin{itemize}
\item  For computation, radix R = 3 would be more economical than R = 2 because the ``optimal'' radix would be R = e = 2.718, according to a demonstration presented  in \cite{Hurst}. It was the motivation for the ternary circuits and more specifically the Russian Setun Ternary Computer developed in 1958 at Moscow State University \cite{Setun}. The same argument can be found in many proposals of ternary circuits. A typical quote is: ``The most efficient multiple-valued system, which leads to the least product cost and complexity, is ternary logic" \cite{Rezaie}.
\item Multivalued circuits  having more logical states, more information could be transmitted through wires, reducing the amount of interconnections inside and outside a chip. This second argument can be found in nearly every proposal of m-valued circuits. We just present one  among a lot of similar quotes: ``One of the main problems in binary logic is the high volume of interconnections which can increase the chip area and power consumption" \cite{Sha}.
\end{itemize}

\begin{figure}[htbp]
\centerline{\includegraphics  [width = 8 cm]{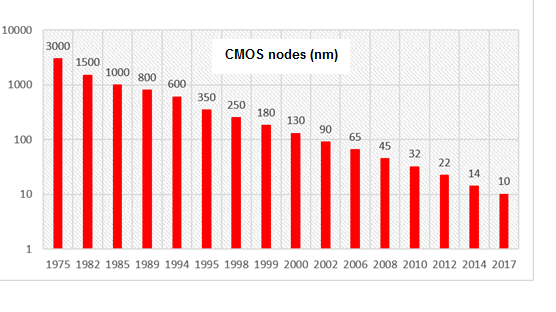}}
\caption{CMOS technological nodes}
\label{figurenode}
\end{figure}

We are faced to a significant contradiction. The ternary radix is supposed to be the most efficient one. However, digital electronics is based on binary circuits. For more than 40 years, the doubling of the number of transistors every N months (12, then 18, then 24) according to Moore's law has been realized by regularly launching a new generation of CMOS technologies, called a technological node. The different successive nodes are shown in Figure \ref{figurenode}. If Moore's law is slowing, 7-nm nodes were already used in 2018 and next ones are announced. 

The interconnection argument raises another question. It is quite evident that ternary operators have less input and output connections than the corresponding binary ones. The reduction factor corresponds to  log(3)/log(2) = 1.58, which is the ratio between information  transmitted by ternary and binary wires. Typical values are shown in Table \ref{Trit/bit}. For 64 bits, there are 36\% less wires with ternary operators. However, the reduced wire count is only applicable for the input and output wires of the considered ternary operators, whatever operator is considered: gate, adder, multiplier, and so on. What about the number of internal wires to interconnect the transistors that are used to implement the gate, adder or multiplier? Similarly, implementing a 64-bit carry propagate adder needs 64 1-bit adders, while a 41-trit adder only needs 41 1-trit adder. However, the 1-trit ternary adder is efficient only if its hardware complexity  is no more than x1.58 the binary adder complexity. This complexity includes the number of transistors, the chip area, the power dissipation, the overall number of interconnects, propagation delays, etc.

\begin{table}
\centering
\caption{Number of ternary and binary wires}
\label{Trit/bit}
\begin{tabular}{|c|c|c|c|c|}
\hline
Number of bits  &8&16&32&64  \\
\hline
 Number of trits  &5&11&21&41  \\
\hline
\end{tabular}
\end{table}

In this paper, we discuss the first argument:  R = e (3) as the optimal one. Then, we compare the relative complexity between ternary and binary operators.

\section{Refuting the Hurst demonstration}
\subsection{The Hurst demonstration \cite{Hurst}}
The number of digits necessary to express a range of N is given by N=$R^d$, where R is the radix and d the number of digits, rounded to the next highest value. It is assumed that the complexity C of the system hardware is proportional to the digit capacity $ R \times d$ where k is a constant.

\begin{equation}
C = k  (R \times d) = k(R \times  \frac {log N}{log R})
\label{eq}
\end{equation}

Differentiating with respect to R shows that R = e for a minimum cost C.

\begin{figure}[htbp]
\centerline{\includegraphics  [width = 8 cm]{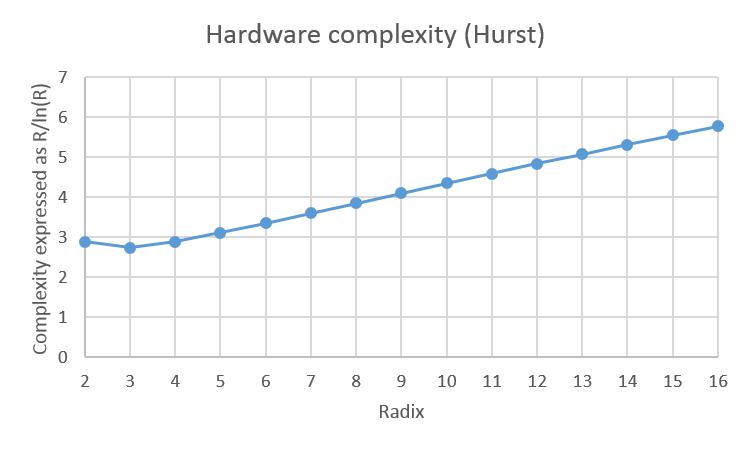}}
\caption{Hardware complexity for Radixes 2 to 16}
\label{H216}
\end{figure}

\begin{figure}[htbp]
\centerline{\includegraphics  [width = 8 cm]{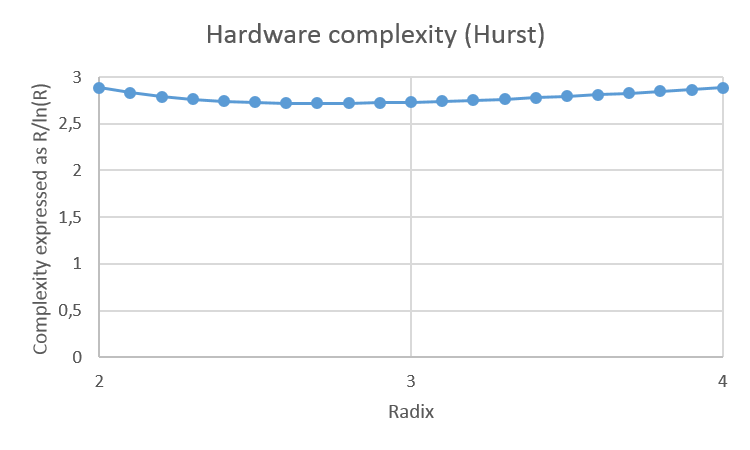}}
\caption{Hardware complexity for Radixes 2 to 4}
\label{H24}
\end{figure}

Fig. \ref{H216} shows that the curve complexity = f(R) for radixes 2 to 16 presents a minimum close to R = 3. Figure \ref{H24} is a zoom of Figure \ref{H216} for R = 2 to 4.

Fig. \ref{H24} leads to two different remarks:
\begin{itemize}
\item  There is a minimum value for R = e, but the curve is very flat. $C(e)=2.718$  while $C(2) =2.885$ and $C(4)=2.885$. The difference between C(2) and C(3) is 5.66\%. Is such a difference sufficient to claim that ternary circuits are more efficient than binary ones?
\item It turns out that C(2) = C(4). Does binary circuits and quaternary circuits carrying the same amount of  information have the same hardware complexity? It is very easy to find 4-valued circuits that are far more complex than the two corresponding binary ones. Such an example is provided below. 
\end{itemize}

\subsection{A comparison of two binary inverters with one 4-valued inverter }
\begin{figure}[htbp]
\centerline{\includegraphics  [width = 8 cm]{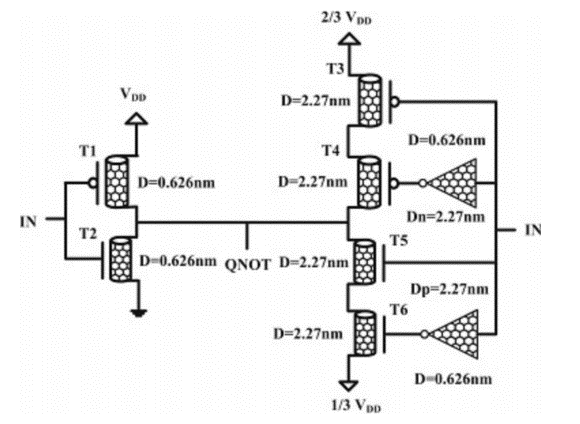}}
\caption{A 4-valued CNTFET inverter}
\label{4VINV}
\end{figure}

Fig. \ref{4VINV} presents a 4-valued CNTFET inverter that has been presented in Microelectronics Journal in 2015 \cite{Navi}. At that point in the discussion, there is no need to give details on the CNTFET technology as we compare transistor counts in the same technology. A 4-valued inverter carries 2 bits of information. Two binary inverters also carry 2 bits of information and use $2 \times 2 = 4$ transistors. The 4-valued inverter has 10 transistors. The 4-valued inverter has 10/4 = x2.5 more transistors. Looking carefully to Fig. \ref{4VINV}, we can observe that  two binary inverters are included in the right part of the figure, which means that transistors T1 to T6 are the overhead of the 4-valued approach over the binary one! More, the 4-valued inverter needs three voltage supplies ($ V_{dd}/3$ , $ 2V_{dd}/3$ and $ V_{dd}$  ) while the binary inverter only uses one. 
Many other examples could be provided with different technologies and different circuit styles. 
This result is not surprising. The binary values 00, 01, 10 and 11 are organized according to the Boolean lattice: each binary inverter has only one threshold level. The quaternary values are totally ordered: $0 < 1 < 2 < 3$ and the 4-valued inverter has 3 threshold levels. The number of threshold levels affects the hardware complexity.

\subsection{Refining Hardware System complexity}

The problem with equation \ref{eq} is that the system cost is only considered proportional to the digit capacity $R \times d$, implicitly assuming that the hardware cost is the same for any value of R, which obviously is not true as previously shown. Let's assume that the hardware complexity is proportional to $R -1$, i.e. the number of threshold levels.   The new equation is:
\begin{equation}
C =  kR(R-1)  \frac {log N}{log R}
\label{eq1}
\end{equation}

\begin{figure}[htbp]
\centerline{\includegraphics  [width = 8 cm]{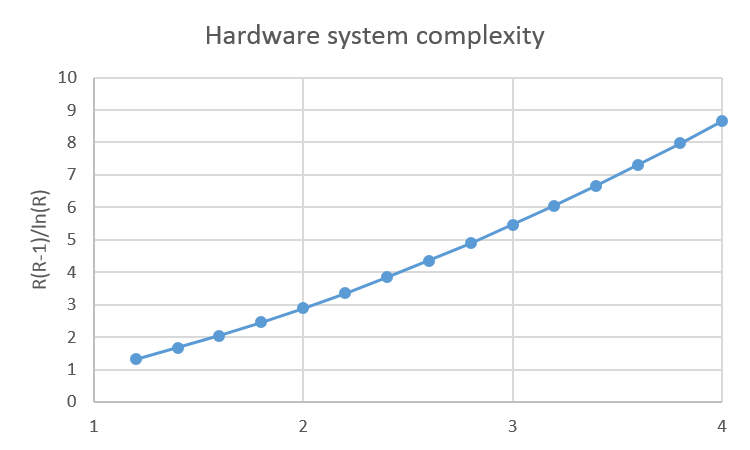}}
\caption{New hardware system complexity}
\label{NHC}
\end{figure}

Fig. \ref{NHC} presents the curve C=f(R) corresponding to Equation \ref{eq1}. Now, there is no minimal value. C=f(R) is continuously rising and the optimal radix is 2. Again, this result is not surprising. It corresponds to what has been observed with the tremendous story of binary integrated circuits for more than five decades.

\section{Comparing ternary and binary circuits in the same technology}
\subsection{Methodology}
Hardware complexity is difficult to define as many parameters can be considered:
\begin{itemize}
\item Number of transistors
\item Number of interconnections
\item Chip area
\item Power dissipation
\item Propagation delays
\item Etc.
\end{itemize}
Obviously, the most significant information is the chip area and power dissipation of fabricated chips in a given technology. However, comparing ternary and binary circuits according to chip area and power dissipation is quite impossible as there are very few or no integrated ternary circuits available for comparisons. 

So comparisons must be done with a simple criterion that is available from the circuit electrical scheme. We use the number of transistors. Although the transistor count is only an estimation, it gives significant insights. In fact, when using the same technology to implement the same operator, it is very doubtful that:
\begin{itemize}
\item More transistors lead to less interconnects
\item More transistors lead to less chip area
\item More transistors lead to less power dissipation
\end{itemize}
When the difference in transistor counts is limited to a few \%, no conclusion can be derived. However, if the transistor count for ternary circuits is x2, x3 or more than for binary circuits when the information ratio  $IR=log(3)/log(2)= 1.58$, it only means that the ternary circuits have more interconnects, more chip area, more power dissipation than the corresponding binary ones and are worthless. Figure \ref{4VINV} was a good example.
\subsection{Using CNTFET technology for comparing ternary and binary circuits}
A carbon nanotube field-effect transistor (CNTFET) refers to a field-effect transistor that uses a single carbon nanotube or an array of carbon nanotubes as the channel material instead of bulk silicon in the traditional MOSFET. The MOSFET-like CNTFETs having p and n types look the most promising ones. This technology has advantages and drawbacks:
\begin{itemize}
\item CNTFET have variable threshold voltages (according to the inverse function of the diameter). Among advantages, high electron mobility, high current density, high transductance can be quoted.
\item Lifetime issues, reliability issues, difficulties in mass production and production costs are quoted as disadvantages.
\end{itemize}
We use this technology for several reasons:
\begin{itemize}
\item This technology is one of the few proposed ones  to overcome the limitations of the FinFET technologies after the end of Moore's law.
\item The MOSFET-like CNTFETs have the same circuit styles than the CMOS technologies, which means that the comparison results are not limited to that technology.
\item A large number of CNTFET ternary or m-valued circuits have been proposed in the recent last years. We will use these proposals for the comparisons.
\end{itemize}

\section {Transistor counts for ternary and binary operators}
The comparison is done at different levels:
\begin{itemize}
\item Gate level: inverters, Nor and Nand gates
\item Arithmetic circuits such as adders and multipliers.
\item Sequential circuits: D latch and D flip-flop
\item Static memory cell
\end{itemize}
At each level, the transistor count ratio is compared with the information ratio (1.58).

\subsection {Inverters and basic gates}

\begin{figure}[htbp]
\centerline{\includegraphics  [width =8 cm]{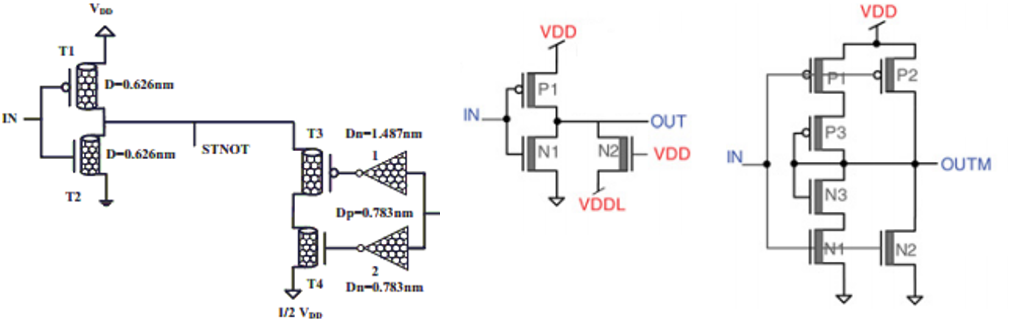}}
\caption{Ternary inverters proposed in \cite{Navi},  \cite{Nepal}, \cite{lin}}
\label{TINV3}
\end{figure}

Ternary inverters and logic gates have three logical levels 0, 1 and 2. Voltage levels 0,$ V_{dd}/2$ and $ V_{dd}$ can be associated with the logical levels. It should be outlined that the voltage levels of the corresponding binary circuits are 0 and $ V_{dd}/2$ as power dissipation requires to use the minimal  power supply voltage. 
It is obvious that 0 and 2 correspond to ground (0) and $ V_{dd}$. For level 1, there are two possibilities: using one more power supply ( $ V_{dd}/2$) or getting a  $ V_{dd}/2$ voltage through a resistor divider. The first option needs one additional  power supply. The second one introduces a static power dissipation as a current flows from $ V_{dd}$ to ground. 
\begin{itemize}
\item Left part of Fig. \ref{TINV3} presents a ternary inverter \cite{Navi} with an additional power supply. The level 1 is implemented by transistors T3, P4 and gates 1 and 2  when T1 and T2 are off.
\item  Right part of Fig. \ref{TINV3} presents a ternary inverter \cite{lin} using a resistor divider. The level 1 is implemented by transistors N1, N3, P1, P3 when N2 and P2 are off. Paper \cite{sahoo} presents a similar CNTFET inverter.
\item Middle part of Fig. \ref{TINV3} presents a ternary inverter \cite{Nepal}. The level 1 is implemented by N2 when N1 and P1 are off. This inverter has only 3 transistors. However, with an additional power supply and an always ON transistor, it  combines the drawbacks of the two previous inverters. 
\end{itemize}
The corresponding binary inverter only uses two transistors.
Table \ref{inverter} gives the transistor count and ratio for ternary and binary inverters.  We can observe that the 3T inverter \cite{Nepal} is the only one for which the 1.5 transistor ratio is smaller than the 1.58 information ratio. The difference is small and don't counterbalance the two quoted drawbacks: additional power supply and dc current flow.

\begin{table}[!b]
\caption{Transistor count for ternary and binary inverters}
\label{inverter}
\begin{tabular}{|c|c|c|c|c|}
\hline
&\cite{Navi}&\cite{lin}&\cite{Nepal}&bin\\
\hline
Transistor count  & 8  & 6&3&2 \\
Transistor ratio   & 4  & 3 &1.5& 1 \\
\hline
\end{tabular}
\end{table}

\begin{figure}[htbp]
\centerline{\includegraphics  [width =8 cm]{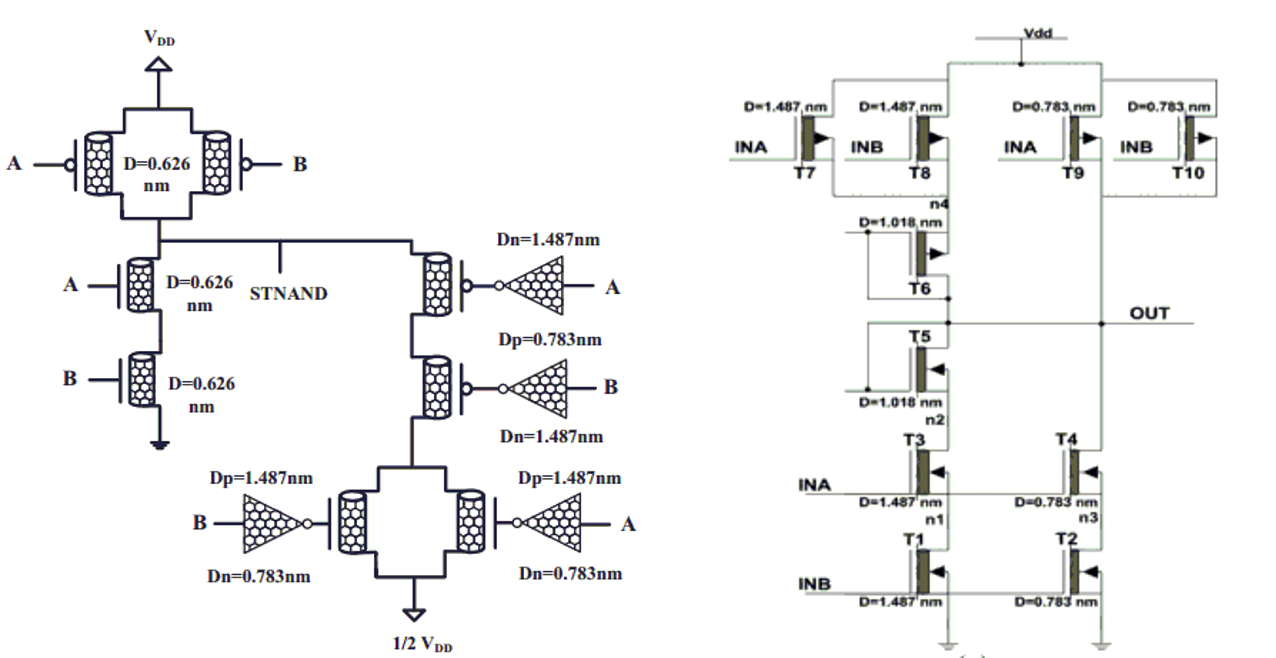}}
\caption{2-input Nand gates proposed in \cite{Navi} and  \cite{lin}}
\label{Nand2}
\end{figure}

Fig. \ref{Nand2} presents the 2-input ternary Nand gates proposed in \cite{Navi} and  \cite{lin} . In both cases, the corresponding 2-input Nand gate has only four transistors. The 2-input ternary Nand corresponding to the 3-transistor inverter would only add the ON transistor to the usual Nand scheme.
Table \ref{Nand} gives the transistor count and ratio for 2-input and N-input ternary and binary Nands. 
Due to the duality between Nand and Nor gates, the results for Nor gates is exactly the same. In the left part of Fig. \ref{Nand2}, we can observe that the ternary Nand gate includes one binary Nand gate, one binary Nor gate + two inverters. In the right part of Fig. \ref{Nand2}, we can observe that the 2-input ternary Nand includes two Nand inverters + the two middle transistors that are needed for the $V_{dd}/2$ level. Only the third approach \cite{Nepal} keeps a transistor count advantage over binary Nands.

\begin{table}
\centering
\caption{Transistor count for ternary and binary Nand gates}
\begin{tabular}{|c|c|c|c|c|c|c|c|cclcl}
  \hline
 Nb inputs&\cite{Navi}&\cite{lin}&\cite{Nepal}&bin&\cite{Navi}/bin&\cite{lin}/bin&\cite{Nepal}/bin\\
  \hline\
2&16&10&5&4&4&2.5&1.25\\
N&6N&4N+2&2N+1&2N&3&2+1/N&1+1/2N\\
  \hline
\end{tabular}
\label {Nand}
\end{table}

\subsection{Adders}
To profit from the reduced number of inputs and outputs, the ternary adder complexity should not be greater than x1.58 the complexity of the binary adder. We consider binary 1-bit and ternary 1-trit adders as they are the building block of any type of adder, from carry-propagate ones to more complicated ones. 
\subsubsection {Binary and Ternary Hall adders}
Unfortunately, ternary half adders cannot be implemented just by using Nand, Nor and inverter gates. While ternary basic logic gates  can be derived from the binary gates by adding circuitry to implement the ``middle" level, this approach can no longer be used when the truth table is more complicated than the basic gate truth tables.
The truth table of a ternary half adder is given in Tab \ref{thruthtable}. 

\begin{table}
\centering
\caption{Truth table of a ternary half adder}
\begin{tabular}{|c|c||c|c|}
  \hline
 a&b&s&cout\\
\hline
 0&0&0&0\\
 0&1&1&0\\
 0&2&2&0\\
 1&0&1&0\\
 1&1&2&0\\
 1&2&0&1\\
 2&0&2&0\\
 2&1&0&1\\
 2&2&1&1\\
  \hline
\end{tabular}
\label {thruthtable}
\end{table}

\begin{figure}[htbp]
\centerline{\includegraphics  [width = 8 cm]{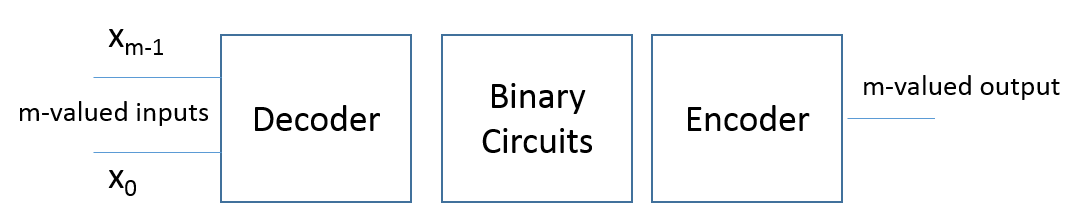}}
\caption{General scheme of m-valued circuits.}
\label{fig}
\end{figure}

\begin{table}
\centering
\caption{Truth table for ternary to binary decoder}
\renewcommand*{\arraystretch}{1.4}
\begin{tabular}{|c|c|c|c|c|c|}
 \hline
A&$\overline{A1}$&A1&$\overline{A0}$&A0\\
\hline
\renewcommand*{\arraystretch}{1}
 0&2&0&2&0\\
1&2&0&0&2\\
2&0&2&0&2\\
  \hline
\end{tabular}
\label {decoder}
\end{table}

\begin{table}
\centering
\caption{Truth table for the binary to ternary encoder}
\renewcommand*{\arraystretch}{1.4}
\begin{tabular}{|c|c||c|}
 \hline
$\overline{Sum1}$&$\overline{Sum2}$&$Sum$\\
\hline
\renewcommand*{\arraystretch}{1}
 2&2&0\\
0&2&1\\
0&0&2\\
  \hline
\end{tabular}
\label {encoder}
\end{table}

To implement an arbitrary truth table, the computation process obeys to the general scheme of m-valued circuits (Fig. \ref{fig}) that was presented in \cite{b1}. The decoder truth table is given in Table \ref{decoder}. The binary outputs of the decoder circuits are 0 and 2. From this table, we derive the following information:
\begin{itemize}
\item $X=0$ when $\overline{X0} =2$.
\item  $X=1$ when $\overline{X1}.X0 =2$.
\item $X=2$ when $X1 =2$.
\end{itemize}

\begin{figure}[htbp]
\centerline{\includegraphics  [width = 8 cm]{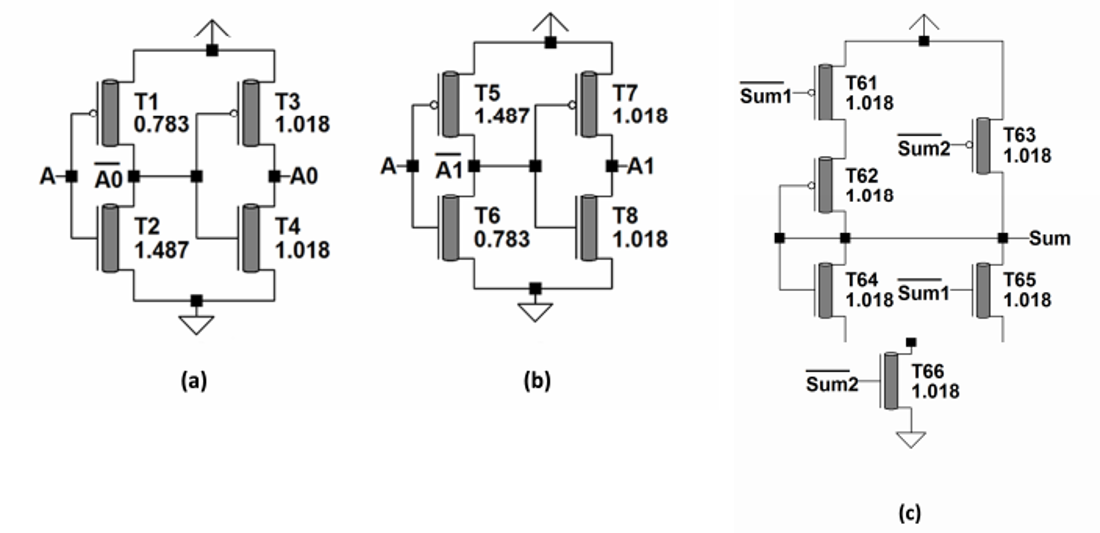}}
\caption{Decoder and encoder circuits \cite{sahoo}.}
\label{decenc}
\end{figure}

In the remaining part of this paper, we consider the approach that is used by references \cite{lin} and \cite{sahoo}. The encoder circuit uses 6 transistors instead of 3 with the 3-transistor inverter approach. While the transistor counts could be slightly greater, the difference is not significant, as the binary part in Fig. \ref{fig} always uses most of the transistors.
The decoder and encoder circuits used in \cite{sahoo} are shown in Fig. \ref{decenc}. Part (a) corresponds to the threshold detector  $V_{dd}/2$ and $V_{dd}$. The CNT diameters of T1 and T2 set up the threshold for the first inverter while the second T3-T4 inverter is a normal binary inverter. Similarly, part (b) present the threshold detection between 0 and $V_{dd}/2$. Part (c) of the figure presents the binary to ternary encoder, which behaviour is shown in Table \ref{encoder}. In this table, the inputs are named $\overline{sum1}$ and $ \overline{sum2}$ as in \cite{sahoo}. 

For the ternary full adder presented in \cite{sahoo}, the binary equations are thus:
\begin{equation}
Sum1=\overline{A0}.\overline{B1}.B0 + \overline{A1}.A0.\overline{B0}+A1.B1
\end{equation}
\begin{equation}
Sum2=\overline{A0}.B1 + \overline{A1}.A0.\overline{B1}.B0+A1.\overline{B0}
\end{equation}
\begin{equation}
Cm=B1.A0 + A1.B0
\end{equation}

\begin{figure}[htbp]
\centerline{\includegraphics  [width = 8 cm]{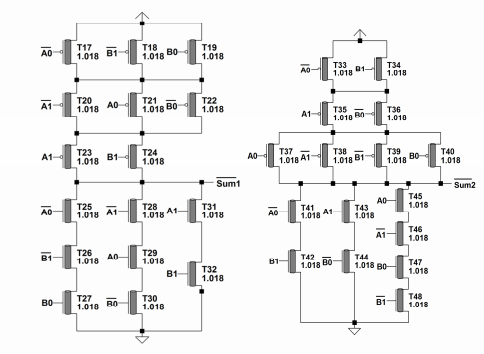}}
\caption{Sum binary parts.}
\label{sum12}
\end{figure}

\begin{figure}[htbp]
\centerline{\includegraphics  [width = 8 cm]{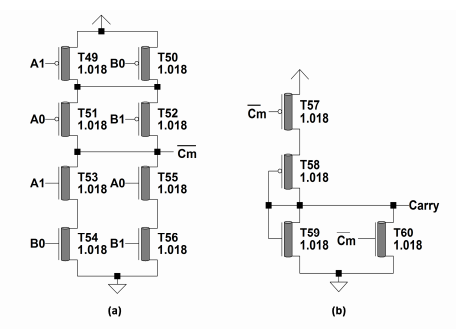}}
\caption{Carry generation.}
\label{carry}
\end{figure}

The corresponding circuit generating $\overline{sum1}$ and  $\overline{sum2}$  is presented in Fig. \ref{sum12}. The carry generation is presented in Fig. \ref{carry}. The generation of the carry output is decomposed into two parts. The first one delivers the binary carry output with levels 0 and $V_{dd}$. However, the ternary carry output has only 0 and 1 logical values. This means that a special carry encoder is needed that is shown in the  part (b) of the figure.

The overall ternary half adder has a total of 16 + 16 + 16 + 6 + 12 = 66 transistors

\begin{figure}[htbp]
\centerline{\includegraphics  [width = 8 cm]{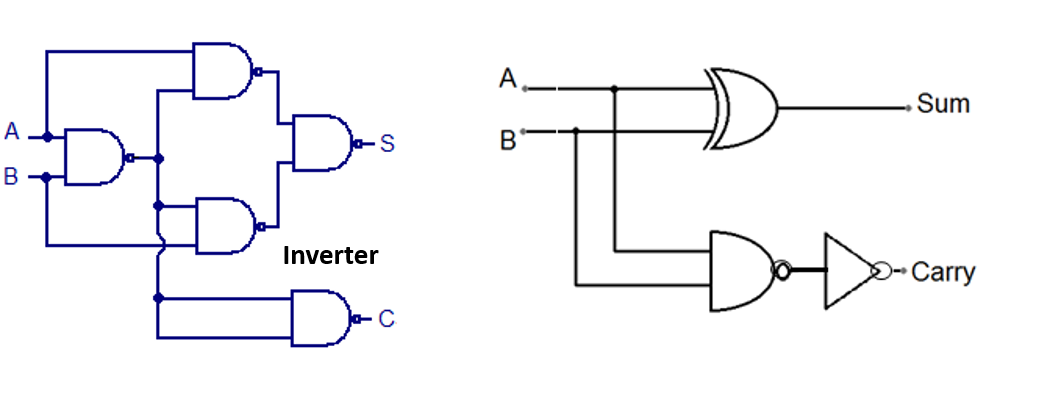}}
\caption{Binary half adders.}
\label{binha}
\end{figure}

\begin{figure}[htbp]
\centerline{\includegraphics  [width = 8 cm]{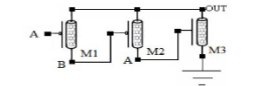}}
\caption{CNTFET 3T Xor}
\label{3Txor}
\end{figure}

There are different possible implementations of a binary half adder. A first one is presented in the left part of  Fig \ref{binha} : it only uses Nand gates and one inverter. A second one is presented in the right part of   Fig \ref{binha} : it is the typical implementation using a Xor gate that can be implemented with 3 transistors as proposed in \cite{nehru} (Fig.\ref{3Txor}). It should be outlined that the 3T Xor like any circuitry using pass transistors doesn't restore levels with the corresponding potential drawbacks: noise margins and propagation delays due to the series of pass transistors.
The transistor counts of the ternary and binary versions are given in Table \ref {HA counts} together with the transistor ratios between ternary and binary versions. 

\begin{table}
\centering
\caption{Transistor counts and ratios for ternary and binary half adders}

\begin{tabular}{|c|c|c|c|}
 \hline

 &3-HA&2-HA without Xor&2- HA with Xor\\
  \hline
Counts&66&18&9\\
Ratio 3/2&1&1/3.67&1/7.3\\
  \hline
\end{tabular}
\label {HA counts}
\end{table}

\subsubsection {Binary and Ternary Full adders}

\begin{table}
\centering
\caption{Truth table of a ternary full adder}
\begin{tabular}{|cc|cc||cc|cc|}
\hline
 \multicolumn{4}{|c||}{cin=0} & \multicolumn{4}{|c|}{cin=1}\\
  \hline
 a&b&s0&cout0&a&b&s1&cout1\\
\hline
 0&0&0&0&0&0&1&0\\
 0&1&1&0&0&1&2&0\\
 0&2&2&0& 0&2&0&1\\
 1&0&1&0&1&0&2&0\\
 1&1&2&0&1&1&0&1\\
 1&2&0&1&1&2&1&1\\
 2&0&2&0&2&0&0&1\\
 2&1&0&1&2&1&1&1\\
 2&2&1&1&2&2&2&1\\
  \hline
\end{tabular}
\label {3TT}
\end{table}

The truth table of a ternary full adder is presented in Table\ref{3TT}. The simpliest way to implement it is to generate sum10 and sum20 when cin=0 (left part of Table \ref {3TT}) and sum11 and sum21 when cin=1 (right part of Table \ref {3TT}). It turns out that
sum21=sum10 (equation 1). The only additional computations are  sum11 and cm1, plus the multiplexers to get $\overline{sum1}$, $\overline{sum2}$ and $\overline{cm}$ according to cin. The different logical equations are:
\begin{equation}
Sum10=Sum21=\overline{A0}.\overline{B1}.B0 + \overline{A1}.A0.\overline{B0}+A1.B1
\end{equation}
\begin{equation}
Sum20=\overline{A0}.B1 + \overline{A1}.A0.\overline{B1}.B0+A1.\overline{B0}
\end{equation}
\begin{equation}
Sum 11=\overline{A0}.\overline{B0} + \overline{A1}.A0.B1 + A1.\overline{B1}+ A1.\overline{B1}.B0
\end{equation}
\begin{equation}
Cm0=B1.A0 + A1.B0
\end{equation}
\begin{equation}
Cm1=A1+ B1 + \overline{A1}.A0 + \overline{B1}.B0
\label{cm1}
\end{equation}
The multiplexers correspond to the following equations:
\begin{equation}
\overline{Sum1}=\overline {\overline{cin}.Sum10 + cin.Sum11 }
\end{equation}
\begin{equation}
\overline{Sum2}=\overline {\overline{cin}.Sum20 + cin.Sum21 }
\end{equation}
\begin{equation}
\overline{cm}=\overline {\overline{cin}.cm0 + cin.cm1 }
\end{equation}

Table \ref{ST} summarizes the number of transistors needed to implement the different functions. The total number of transistors of the ternary full adder is thus 66+ 58 = 124.

\begin{table}
\centering
\caption{Number of of additional transistors from half adder to full adder}
\begin{tabular}{|c|c||c|}

  \hline
function&gate&transistor count\\
\hline
$\overline{cin}$&inverter&2\\
sum11&complex gate&20\\
cm1&complex gate& 12\\
$\overline{ sum1}$&complex gate&8\\
$\overline{ sum2}$&complex gate&8\\
$\overline{cm}$&complex gate&8\\
\hline
total&&58\\
  \hline
\end{tabular}
\label {ST}
\end{table}

\begin{figure}[htbp]
\centerline{\includegraphics  [width = 8 cm]{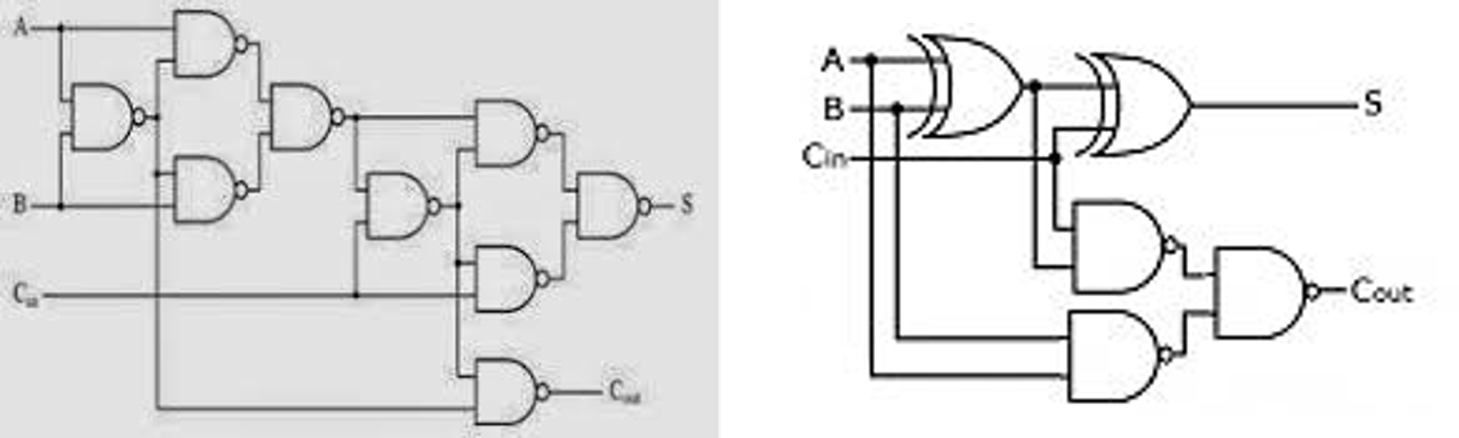}}
\caption{Binary full adders}
\label{FA}
\end{figure}

\begin{figure}[htbp]
\centerline{\includegraphics  [width = 9 cm]{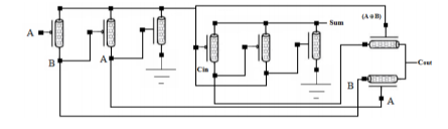}}
\caption{8T binary full adder}
\label{8TFA}
\end{figure}

Fig \ref{FA} presents two typical implementations of a full adder. The left only uses Nand gates. The right part uses Xor and Nand gates. A CNTFET 8T full adder (Fig \ref{8TFA}) has been presented \cite{nehru}. Again, this adder doesn't restore levels and using it could raise issues. The transistor counts of the ternary and binary versions are given in  Table \ref {FA counts} together with the transistor ratio between ternary and binary versions. 

\begin{table}
\centering
\caption{Transistor counts and ratios for ternary and binary full adders}

\begin{tabular}{|c|c|c|c|c|}
 \hline

 &3-FA&Nand 2-FA&Xor 2-FA&8T 2-FA\\
  \hline
Counts&124&36&18&8\\
Ratio 3/2&1&1/3.45&1/6.9&1/15.5\\
  \hline
\end{tabular}
\label {FA counts}
\end{table}

Obviously, the estimation of propagation delays is more difficult without simulations. A rough comparison can be done.
For the most conservative version of the binary adder (left part of Fig. \ref{FA}, the critical path is a series of 6 2-input Nand gates. Except for the last gate, every Nand fan-out is 1. For the ternary adder, the critical path is a series of the following items:
\begin{itemize}
\item 2 inverters for the decoder. The outputs of the decoder are heavily loaded by the different complex gates ($\overline{sum10},\overline{sum20},\overline{sum11},\overline{cm0}$ and $\overline{cm1}$).
\item a complex gate :$ \overline{sum10}$ or $\overline{sum20}$  or $\overline{sum11}$or $\overline{cm1}$. With a large number of transistors, any of these complex gates has a propagation delay far greater than a 2-input Nand gate.
\item a complex gate that implements a multiplexer: $ \overline{sum1}$ or $\overline{sum2}$ or $ \overline{cm}$
\item the sum or carry final encoder circuits. The sum encoder is probably the critical one.
\end{itemize}
It is pretty sure that the ternary full adder propagation delay is significantly greater than the propagation delay of any version of the binary full adders.

Even with the most conservative implementation of the 1-bit adder, the 1-trit adder has more than x1.58 transistors than the 1-bit adder. A P-trit carry propagate adder is thus more complex that a N-bit carry propagate adder with $ P \geq N/1.58$.

\subsection{Ternary and Binary Multipliers}
The truth table of a 1-trit multiplier is given in Table \ref{1TM}. 

\begin{table}
\centering
\caption{Truth table of a one trit multiplier}
\begin{tabular}{|c|c||c|c|}
  \hline
 a&b&s&cout\\
\hline
 0&0&0&0\\
 0&1&0&0\\
 0&2&0&0\\
 1&0&0&0\\
 1&1&1&0\\
 1&2&2&0\\
 2&0&0&0\\
 2&1&2&0\\
 2&2&1&1\\
  \hline
\end{tabular}
\label {1TM}
\end{table}

Using the same approach as for the adder,
the equations are:
\begin{equation} 
S2= A1.\overline{B1}.B0 + B1.\overline{A1}.A0
\end {equation}
\begin{equation} 
S1= A1.B1 +\overline{B1}.B0.\overline{A1}.A0
\end {equation}
\begin{equation}
Cm=A1.B1
\end{equation}
The number of transistors is 4 (decoder)+ 12 (S1) + 12 (S2) + 6 (s encoder) + 4 (cout encoder) = 38. 

The binary 1-bit multiplier is a And gate, which means 6 transistors (2-nput nand gate + inverter). The ratio is 6.3. One can observe that the 1-bit multiplier has 2 input and 1 output, while the 1-trit multiplier has 2 inputs and \textbf{2 outputs}. This is a curious situation as m-valued logic is supposed to reduce the number of interconnections!

As for adders, comparing N-bit binary and P-trit ternary multipliers is interesting. We only make the comparison for N=8 and P=5.
As a matter of fact, $2^8=256$ while $3^5 = 243$, which means that the ternary multiplier has slightly less computational capability than the binary one.
The comparison includes two parts:
\begin{itemize}
\item the set of 1-bit and 1-trit multipliers. 
\item the tree to reduce the set of partial products into a final sum of two reduced partial products. The Wallace tree is one of the possible reduction tree.
\end{itemize}

A 8x8 bit multiplier uses $8^2 = 64 $  1-bit multiplier while a 5x5 trit multiplier only use $5^2=25$ 1-trit multipliers. The ratio is $64/25 = 2.56$. Obviously, this value is close to $(\frac{Log 3}{Log2})^2$. However, the binary multiplier generates 8 partial products, while the ternary multiplier generates 10 partial products, 5 ternary ones and 5 binary ones provided by the ternary output and the binary carry output of each 1-trit multiplier. To benefit from the smaller number of 1-trit multiplier, its complexity should not be greater than x2.56 the complexity of the 1-bit multiplier, which is not the case with the 6.3 ratio that was computed.

\begin{figure}[htbp]
\centerline{\includegraphics  [width = 8 cm]{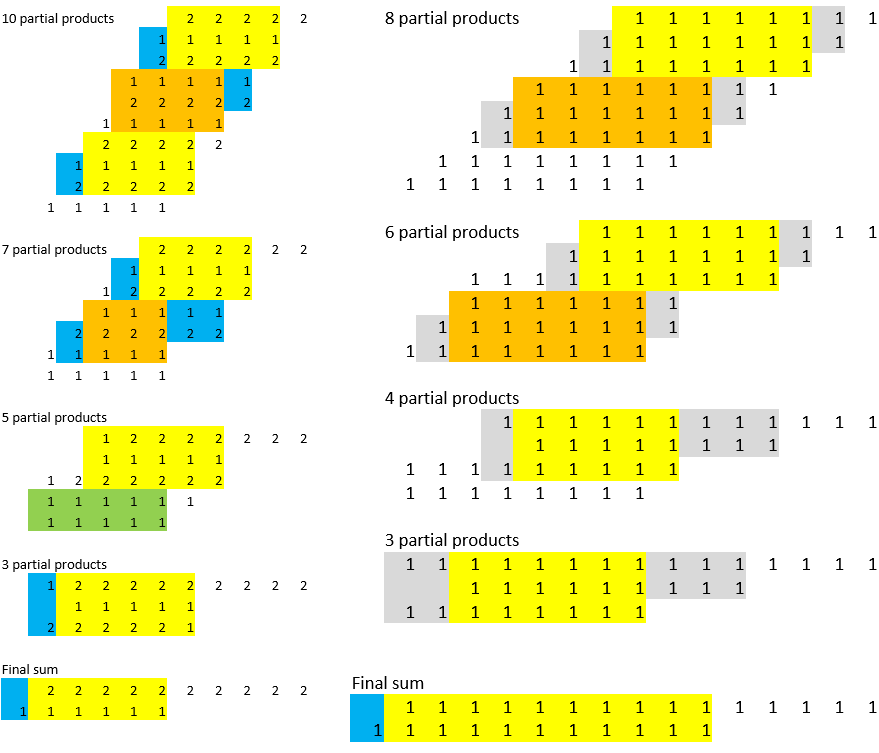}}
\caption{Wallace trees for 5x5 ternary (left) and 8x8 binary (right) multipliers}
\label{W85}
\end{figure}

 Fig. \ref{W85} compares the Wallace trees for the ternary and the binary multiplies. The left part presents the 5-stage ternary tree. 3 and 2 corresponds to the ternary and binary values. Ternary half  and full adders are respectively showed with blue and yellow colors. The right part presents the 4-stage binary tree. Binary half and full adders also use blue and yellow colors.

The ternary Wallace tree uses 35 T-FAs and 7 T-HAs. When using a carry-propagate adder (CPA) for the final addition, the overall count is 38 T-FAs and 8 T-HAs. The binary Wallace tree uses 38 B-FAs and 15 B-HAs. When using CPA for the final addition,the overall count is 47 B-FAs and 17 B-HAs. So, the binary multiplier use 1.24 x more FAs and 2.12 x more HAs. However these ratios are too small to compensate the huge complexity disadvantage of the T-FAs and T-HAs compared to the binary ones. 

The 5x5 ternary multiplier uses 6190 transistors versus 2382 for the 8x8 binary ones, i.e. x2.6 more transistors.

\begin{table}
\centering
\caption{Transistor counts for 8x8 binary multiplier versus 5x5 ternary one}
\begin{tabular}{|c|c|c|c||c|c|c||c|}
  \hline
Radix&1-digit X&HA&FA&$\Sigma{ 1digit X}$&$\Sigma HA$&$\Sigma FA$&Total\\
 \hline
Binary&6&18&36&	384&306&1692&2382\\
Ternary&38&66&124&950&528&4712&6190\\
 \hline
\end{tabular}
\label {BTM}
\end{table}

\subsection{Flip Flops and Memory Cells}
\begin{figure}[htbp]
\centerline{\includegraphics  [width = 8 cm]{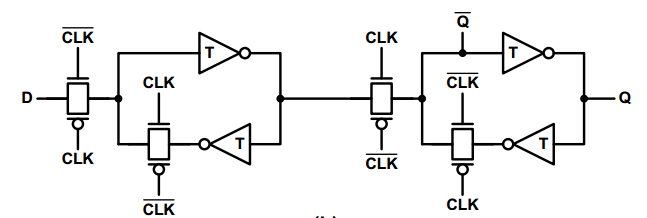}}
\caption{Ternary D-Flip-Flop}
\label{TDFF}
\end{figure}

The D-Flip is the most important flip-flop from which any other one could be derived. Fig. \ref{TDFF} presents a typical ternary D master slave flip flop. It corresponds directly to the binary D Flip Flop by replacing the binary inverters by ternary inverters.
Table \ref{FF} gives the transistor counts for the different ternary and binary D Flip-Flops. Again, only the 3-transistor inverter approach has a transistor ratio less than the 1.58 information ratio. 

\begin{table}
\centering
\caption{Comparison of Ternary and Binary D-Flip Flops}
\begin{tabular}{|c|c|c|c|c|c|}
  \hline
 &\cite{Navi}&\cite{lin}&\cite{Nepal}&bin\\
  \hline\
Trans. count&40&32&20&16\\
Trans. ratio&2.5&2&1.25&1\\
  \hline
\end{tabular}
\label {FF}
\end{table}

\begin{figure}[htbp]
\centerline{\includegraphics  [width = 8 cm]{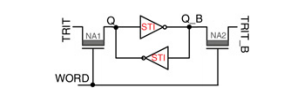}}
\caption{Ternary Memory Cell from \cite{Nepal}}
\label{Tcell2}
\end{figure}

SRAMs use static cells. There are several possible implementations of the memory cell. We consider the simplest ternary cell derived from the classical 6-T binary cell (Fig. \ref{Tcell2}).  Table \ref{SRAM} gives the transistor counts for the different ternary and binary SRAM cells. Again, only the 3-transistor inverter approach has a transistor ratio less than the 1.58 information ratio.

\begin{table}
\centering
\caption{Comparison of Ternary and Binary SRAM Cells}
\begin{tabular}{|c|c|c|c|c|}
  \hline
 &\cite{lin}&\cite{Nepal}&bin\\
  \hline\
Trans. count&14&8&6\\
Trans. ratio&2.33&1.33&1\\
  \hline
\end{tabular}
\label {SRAM}
\end{table}

While Flip-Flops and Static Rams are based on the logic gates that are used to implement combinational operators, Dynamic Rams are based on electrical features instead of logical ones. Binary DRAMs use electrical charges to store bits in 1.5 T cells (1 transistor + 1 capacitance). 
M-valued SRAMs and DRAMs  have been fabricated and tested by industrial companies in the 80's and the 90's, such as Hitachi \cite{Aoki}, NEC \cite{Sugi} and \cite{Murotani}, etc. They are detailed in \cite{Gulak}. During the following decades, there was no longer such presentations by industrial companies. 

M-Valued flash memories also use similar techniques and were presented in the 90s \cite{Bauer,Jung}. They are now largely used.  4-valued (MLC) flash memories store two bits per cell. 8-valued (TLC) memories store 3 bits per cell. In 2018,  ADATA, Intel, Micron, and Samsung have launched some SSD products using QLD NAND-memory with 4 bits per cell. While binary flash memories have the advantage of faster write speeds, lower power consumption and higher cell endurance, M-valued flash memories provide higher data density and lower costs. These M-valued circuits (M=$2^n$) are used for higher density, not for higher speeds. 

\section{Conclusion}
While the tremendous story of binary circuits for more than five decades is the best refutation of the supposed advantage of ternary circuits over binary ones, we have provided a refutation of the demonstration e=2.718 as the best radix for computation. 

Ternary circuits have a log(3)/log(2) = 1.58  information advantage over binary circuits. However, ternary circuits would be interested if (and only if) their complexity in terms of gate count, interconnection count, chip area, propagation delay, etc. is no more than x1.58 the complexity of the corresponding binary circuits. Using MOSFET-like CNTFET circuits for comparing binary and ternary circuits, we have shown that this condition is fulfilled only in one specific case with two conditions: 
\begin{itemize}
\item when implementing the ``middle" level by an ON transistor \cite{Nepal}. This approach uses two power supplies instead of one, and is electrical debatable: there is a dc path and a fight between two transistors for the lower level.
\item when implementing the basic gates (inverter, nand, nor, etc.) or the circuits based on the ternary inverter such as the D Flip-Flop and the SRAM cell.
\end{itemize}
For all the other ternary implementations and for the typical arithmetic circuits such as adders and multipliers, the ternary circuits are outperformed by the binary ones. Ternary circuits cannot avoid using two debatable techniques: either using one additional power supply adding to interconnection issues, or generating the ``middle" level via a dc current flow which introduces static power dissipation.

It does not mean that ternary circuits can never be used: they can be used when they implement some functions having 3  different states. A good example is the ternary content-addressable memories (CAM) which cells store three different states: 0, 1 and X (don't care). It allows to search for words having unknown and non-significant bits. However, faced to the still continuing advances of binary circuits, ternary circuits have been and will probably remain restricted to a small niche \cite{b2}.

\end{document}